\documentclass[9pt,twocolumn,twoside]{osajnl}

\journal{ol} 

\setboolean{shortarticle}{true} 

\title{Asymmetric dual-loop feedback to suppress spurious tones and reduce timing jitter in self-mode-locked quantum dash lasers emitting at 1.55 $\mu$m}

\author [*]{Haroon Asghar} 
\author {John.~G.~McInerney}
\affil{Department of Physics and Tyndall National Institute, University College Cork, Cork, Ireland}
\affil[*]{Corresponding author: haroon.asghar@ucc.ie}
\ociscodes{(140.4050) Mode-locked lasers; (140.5960) Semiconductor lasers; (270.2500) Fluctuations, relaxations, and noise}
\begin{abstract}
 We demonstrate an asymmetric dual-loop feedback method to suppress external cavity side-modes induced in self-mode-locked quantum-dash lasers with conventional single- and dual-loop feedback. In this letter, we report optimal suppression of spurious tones by optimizing the delay in the second loop. We observe that asymmetric dual-loop feedback, with large ($\sim$ 8x) disparity in loop lengths, gives significant suppression in external-cavity side-modes and produces flat RF spectra close to the main peak, with low timing jitter compared to single-loop feedback. Significant reduction in RF linewidth and timing jitter was produced by optimizing delay time in the second feedback loop. Experimental results based on this feedback configuration validate predictions of recently published numerical simulations. This asymmetric dual-loop feedback scheme provides simple, efficient and cost effective stabilization of optoelectronic oscillators based on mode-locked lasers.
\end{abstract}
\begin{document}
\maketitle
{Semiconductor mode-locked diode lasers (MLLs) are compact, rugged and efficient sources of ultra-short, intense, and high repetition frequency optical pulses with many applications such as, all-optical clock recovery, Lidar, optical frequency combs, and telecommunications [1-3]. A major limitation of MLLs for most practical applications is their very high timing jitter and phase noise, as spontaneous emission noise and cavity losses make MLLs prone to broad linewidths and therefore substantial phase noise [4]. To improve timing jitter, several experimental methods such as single cavity feedback [5-8], coupled optoelectronic oscillators (OEOs) [9], injection locking [10-12] and dual-loop feedback [13-15] have been proposed and demonstrated. Of the stabilization techniques demonstrated to date, optical feedback is a promising approach in which an additional reflector creates a compound cavity with a high quality factor, with no need for an external RF or optical source. Due to the existence of the extra mirror, side-bands resonant with the round trip time of the external cavity are generated which affect the overall timing jitter and quality of the RF spectra. To overcome these issues, optoelectronic feedback [9] can also be utilized to stabilize timing jitter and to suppress cavity side-modes by conversion of the optical oscillation (using a fast photodetector) to an electrical signal used in a long feedback loop. This technique does not require an RF source, but requires optical-to-electrical conversion. Recently, a simpler dual-loop feedback technique [13-15] without optical/electrical conversion has been demonstrated to improve timing jitter of the MLLs and to filter or suppress the unwanted spurious side-bands. Dual-loop configurations proposed to date [13, 14] yields sub-kHz linewidth but produces additional noise peaks at frequencies resonant with the inverse of the delay time in the second cavity. This is undesirable in many applications where low noise and flat spectra are required, as in frequency comb generation. Recently the influence of the second feedback delay on  side-mode suppression [16] and timing jitter [17] has been studied numerically. In this letter, we report experimental investigation to eliminate these adverse dynamical effects using asymmetric dual-loop feedback by appropriately choosing the length of second feedback cavity. Best side-mode suppression and lower timing jitter relative to single loop feedback is achieved with the length ratio between the two cavities $\sim$ 8x. It was further observed that RF linewidth and integrated timing jitter were reduced by increasing the length of the second cavity. Our findings suggest that noise stabilization and side-mode suppression depends strongly on additional feedback delay times.}\par
{Devices under investigation are two-section InAs/InP quantum dash mode-locked lasers (QDash MLLs) whose active layers have 9 InAs quantum dash monolayers grown by gas source molecular beam epitaxy (GSMBE) embedded within two barriers and separate confinement heterostructure (SCH) layers (dash in a barrier structure). Both barriers and SCHs consisted of $In_{0.8}Ga_{0.2}As_{0.4}P_{0.6}$ quaternary materials with \(\lambda_{g}\) = 1.55 $\mu$m [18]. Total cavity length was 2030 $\mu$m with absorber lengths 240 $\mu$m (length ratio $\sim$ 11.8\%), giving pulse repetition frequency $\sim$ 20.7 GHz (\(I_{Gain}\) = 300 mA)  and average free-space output powers of a few mW. Mode-locking was obtained without reverse bias applied to the absorber section, and the heat sink temperature was fixed at \(19^0\)C. This is a two-section device but works similarly to a single section self-mode-locked laser, since the absorber is not biased; in this case the amount of minimal residual absorption does not affect the mode-locking mechanism [12]. The absorber and gain sections were isolated by a resistance 9 k$\Omega$. The QDash MLL was mounted p-side up on AlN submounts and copper blocks with active temperature control and electrical contacts formed by wire-bonding.}\par 
\begin{figure}[htbp]
\centering
\includegraphics[width=\linewidth]{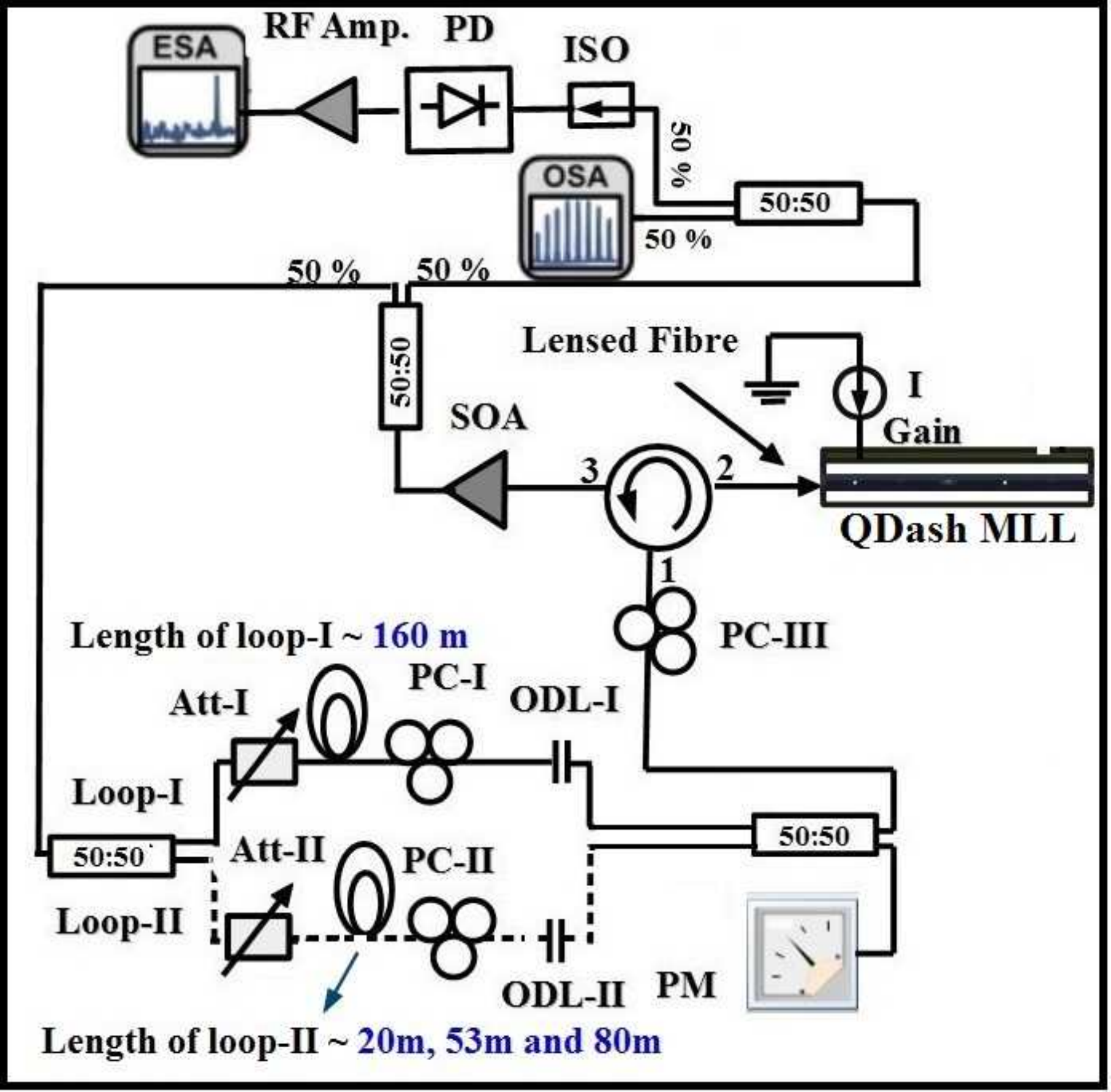}
\caption{Schematic of the experimental arrangement for single (excluding dashed portion) and dual-loop configurations (with dashed portion). \textit{Acronyms}-- SOA: Semiconductor Optical Amplifier; ISO: Optical isolator; PD: Photodiode; ODL: Optical delay line; Att: Optical attenuator; PC: Polarization controller; ESA: Electrical spectrum analyzer; OSA: Optical spectrum analyzer; PM: Power Meter; QDash MLL: Quantum dash mode-locked laser}
\label{fig:false-color}
\end{figure}

 {A schematic for the dual-loop technique is depicted in Fig. 1. For single and dual-loop feedback configurations, a calibrated fraction of light was fed back through port 1 of an optical circulator, then injected into the laser cavity via port 2. Optical coupling loss from port 2 to port 3 was -0.64 dB. The output of the circulator was sent to a semiconductor optical amplifier (SOA) with gain of 9.8 dB, then split into two arms by a 50/50 coupler. 50\% went to an RF spectrum analyzer (Keysight E-series, E4407B) via a 21 GHz photodiode and to optical spectrum analyzers (Ando AQ6317B and Advantest Q8384). The other 50\% of power was split into two equal parts by a 3-dB splitter. For single loop feedback, all power passed through loop-I. For dual-loop configurations (feedback loops-I and-II) the power was split into two loops at the 3-dB splitter. Feedback strengths in both loops were controlled by variable optical attenuators and monitored using power meters. In this experimental arrangement, the length of loop-I was fixed to 160m while the length of second feedback loop was varied in three chosen lengths: 20, 53 and 80m. Polarization controllers in each loop plus one polarization controller before port 1 of the circulator ensured the light fed back through both loops matched the emitted light polarizations to maximize feedback effectiveness. In this experiment the feedback ratio into gain section was limited to $\sim$ -22 dB.}\par
 { {We calculated RMS timing jitter from single sideband (SSB) phase noise spectra measured at the fundamental RF pulse repitition frequency (20.7 GHz) using:
  \begin{equation}
\sigma_{RMS}=\frac{1}{2 \pi f_{ML}}\sqrt{2 \int_{f_{d}}^{f_{u}} L(f)\,df}
\end{equation}where \(f_{ML}\) is the pulse repetition rate, \(f_{u}\) and \(f_{d}\) are the upper and lower integration limits. \textit{L(f)} is the single sideband (SSB) phase noise spectrum, normalized to the carrier power per Hz. To measure RMS timing jitter of the laser in more detail, single-sideband (SSB) noise spectra for the fundamental harmonic repetition frequency were measured. To assess this, RF spectra at several spans around the repetition frequency were measured from small (finest) to large (coarse) resolution bandwidths. The corresponding ranges for frequency offsets were then extracted from each spectrum and superimposed to obtain SSB spectra normalized for power and per unit frequency bandwidth. The higher frequency bound was set to 100 MHz (instrument limited).}\par 
   {To observe the RF spectrum with single-loop feedback, the length of the loop was initially set at 160m; optimally stable resonance occurred when the feedback length was fine tuned using an optical delay line (ODL-I) (which spanned 0 to 84 ps in steps of 1.67 ps). Such optimization provides a resonant condition (at delay setting=13 ps) under which the RF linewidth was reduced from 100 kHz free-running to as low as 4 kHz, with integrated timing jitter to 0.7 ps from 3.9 ps [integrated from 10 kHz - 100 MHz]. Measured phase noise traces as functions of frequency offset from fundamental mode-locked frequency and RF spectra for free running laser (green line) and single loop feedback (gray line) are given in Fig. 5 and Fig. 6, respectively. Under similar delay settings, external cavity side-modes appear in the RF spectrum with frequency spacing 1.28 MHz, the inverse of the loop round trip delay. RF spectra are shown in Fig. 2(a) (gray line) and (b) (gray line), using spans of 10 MHz and 100 MHz, respectively. Frequency resonances can be seen in both frequency spans which contribute significantly to timing jitter, particularly for the longer feedback cavities as they are closer to the main peak and are less suppressed [11]. To eliminate these fluctuations and to improve the side-mode suppression ratio (SMSR), dual-loop feedback was implemented as described in the next section.
   }\par 
   \begin{figure}[htbp]
\centering
\includegraphics[width=\linewidth]{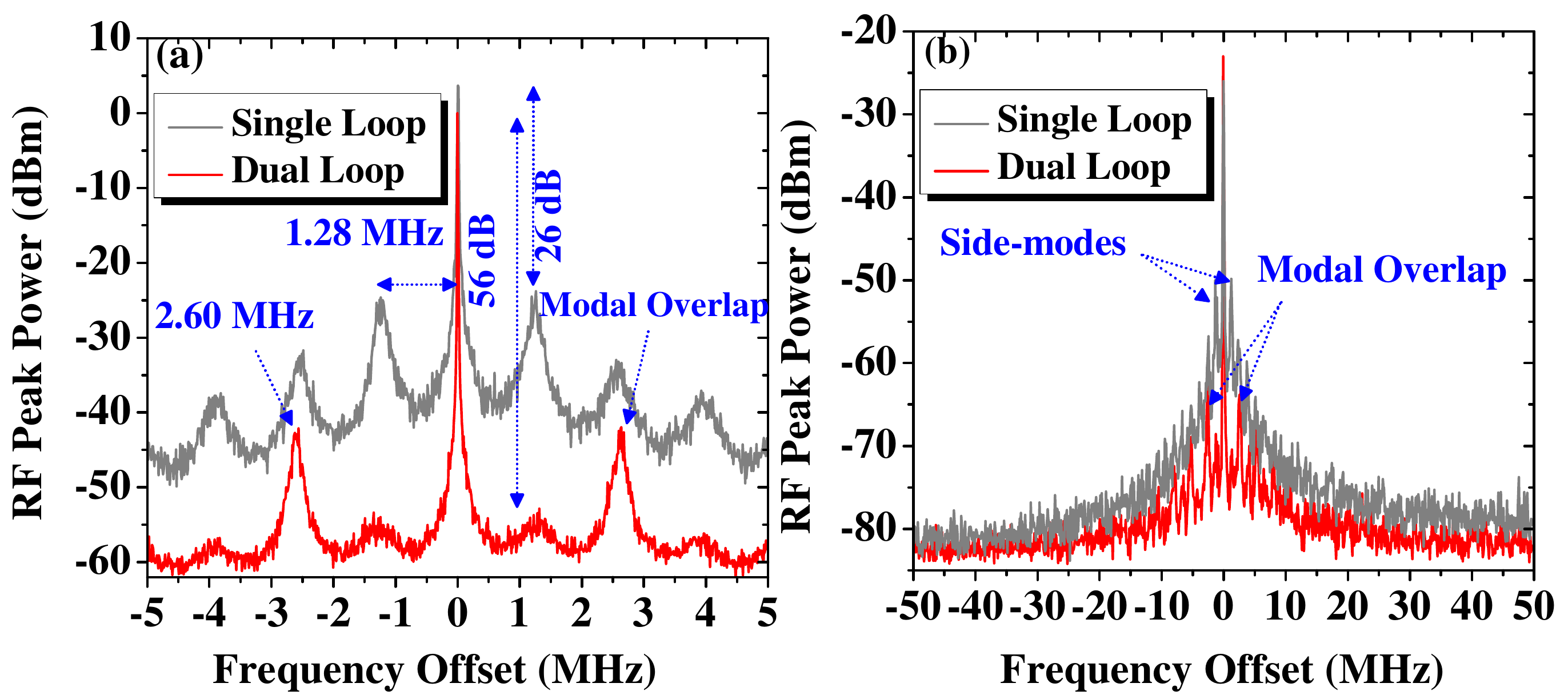}
\caption{RF Spectra of single-loop feedback of length 160m (gray line) and  dual-loops having lengths 160m for loop-I and 80m for loop-II (red line) using frequency span (a) 10 MHz and (b) 100 MHz}
\label{fig:false-color}
\end{figure}
 {To assess the suppression of these frequency resonances, a shorter feedback cavity corresponding to half the period of noise-induced oscillations of loop-I was introduced. Feedback strengths of both cavities were equalized using variable optical attenuators (Att-I and Att-II) plus polarization controllers (PC-I and PC-II) in both loops. One optical delay (ODL-II) was adjusted to full resonance and the length of the other delay line (ODL-I) was tuned over the maximum range available 0-84 ps. When the optical delay lines ODL-I and ODL-II were fine tuned (ODL-I=15 ps and ODL-II=25 ps) so that every second mode of loop-I coincided precisely with a mode of loop-II, maximum >30 dB suppression in the first order side-mode was achieved. However, additional fluctuations (modal overlaps) appeared at frequencies resonant with the inverse of the length of second delay time which becomes the carrier signal. These noise fluctuations depends on the ratio of the loop lengths. Here these fluctuations at frequency spacing 2.60 MHz, are consistent with the length of the second feedback loop 80m. RF spectra of the asymmetric dual-loop configuration are shown in Fig. 2(a) (red line) and (b) (red line), using spans 10 MHz and 100 MHz, respectively. In this fully resonant configuration, the RF linewidth narrowed to < 1 kHz (instrument limited), with timing jitter reduced to 295 fs. Phase noise trace and RF spectra are shown in the Fig. 5 (blue line) and Fig. 6 (blue line), respectively. The RF spectra illustrated in Fig. 2(a) and (b) show that this feedback configuration is not suitable to achieve effective suppression in frequency resonances, as the second delay time will be resonant with the second mode of the first feedback loop which restrict many practical applications where flat and side-band free RF spectra are required. To improve on this situation, a different dual-loop feedback configuration with non-resonant shorter second loop (53m) was investigated, described in the next section.}\par
 \begin{figure}[htbp]
\centering
\includegraphics[width=\linewidth]{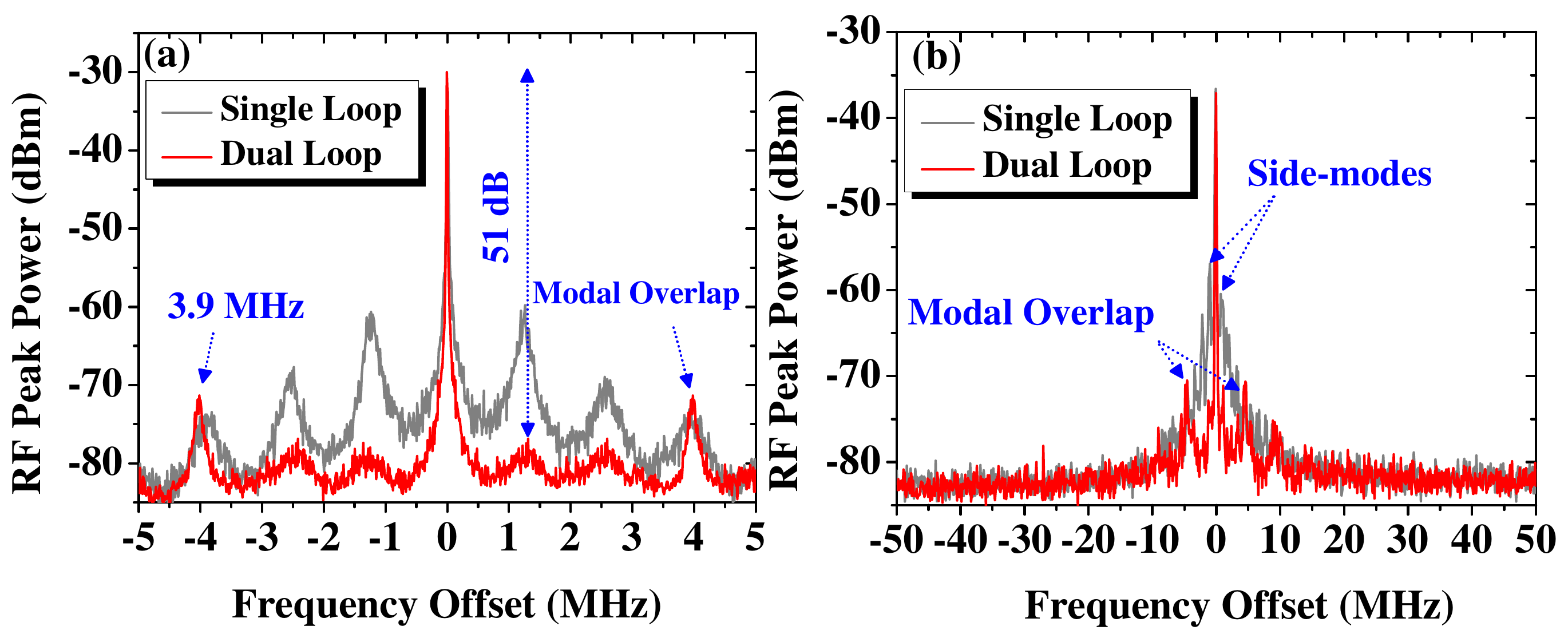}
\caption{RF Spectra of single-loop feedback with length 160m (gray line) and dual-loops having lengths 160m for loop-I and 53m for loop-II (red line) using frequency span (a) 10 MHz and (b) 100 MHz}
\label{fig:false-color}
\end{figure}
{In the dual-loop configuration presented in this section, the length of loop-I was initially set to 160m while that of loop-II was 53m. Upon fine tuning of both optical delay lines (ODL-I=13 ps and ODL-II=15 ps), when the second delay time was resonant with the third harmonic of the first loop, suppression of the first two frequency resonances occurred, while the third harmonic (modal overlap) was unsuppressed. This harmonic was observed at frequency offset 3.9 MHz, corresponding to the 53m length of the outer feedback loop. RF spectra for the asymmetric dual-loop configuration are shown in Fig. 3(a) (red line) and (b) (red line), using spans 10 MHz and 100 MHz, respectively. In this  configuration, when both external cavities are fully resonant, the RF linewidth narrows to 2 kHz with integrated timing jitter 0.45 ps. Phase noise trace and RF spectra are shown in Fig. 5 (black line) and Fig. 6 (black line) respectively. These experimentally measured results show that external cavity side-modes cannot be optimally suppressed by simply choosing the second feedback delay time to be a fraction of the first. To achieve stable and flat RF spectra, we designed an asymmetric dual-loop feedback configuration for effective suppression of external cavity side-modes. Which produced flat RF spectra close to the main peak compared to conventional single- and dual-loop feedback.}\par
{In this asymmetric dual-loop feedback configuration the length of loop-I was fixed (160m) and  loop-II was set $\sim$ 8x shorter than loop-I. Fine tuning of both cavities (ODL-I=15 ps and ODL-II=21 ps) produced precise coincidence of every eighth mode of loop-I with a mode of loop-II, so that strong side-mode suppression occurred and all feedback-induced side-modes and spectral resonances were eliminated under frequency span 10 MHz. RF spectra for this dual-loop feedback configuration (red line) are shown in Fig. 4(a) and (b) with frequency spans 10 MHz and 100 MHz respectively. Furthermore, when both external cavities are fully resonant, the RF linewidth narrows to 8 kHz with integrated timing jitter 0.6 ps. The phase noise trace and RF spectra are shown in Fig. 5 (red line) and Fig. 6 (red line) respectively. In this configuration, the RF linewidth was higher than for single loop feedback, but measured timing jitter was lower; this is due to suppression of external cavity side-modes. In addition, weak modal overlap (with intensity $\sim$ -6 dBm) in RF spectra of dual-loop feedback was noticed at 10.2 MHz frequency spacing, consistent with our 20m outer loop; this is shown in Fig. 4(b)(red line). This behavior shows that effective suppression of external cavity side-modes and reduced timing jitter can be achieved by appropriately fine-tuning the length of the second feedback loop. It should be noted that length of loop-II ($\sim$ 20m) is only optimal in our specific experimental setup. Further reduction in the length of second feedback loop is not possible, as the combined variable optical attenuator, optical delay line, polarization controller and 3-dB coupler have minimum length of $\sim$ 20m. Better suppression of external cavity side-modes could be achieved in an arrangement not subject to this limitation, such as photonic integrated circuit.}\par
\begin{figure}[htbp]
\centering
\includegraphics[width=\linewidth]{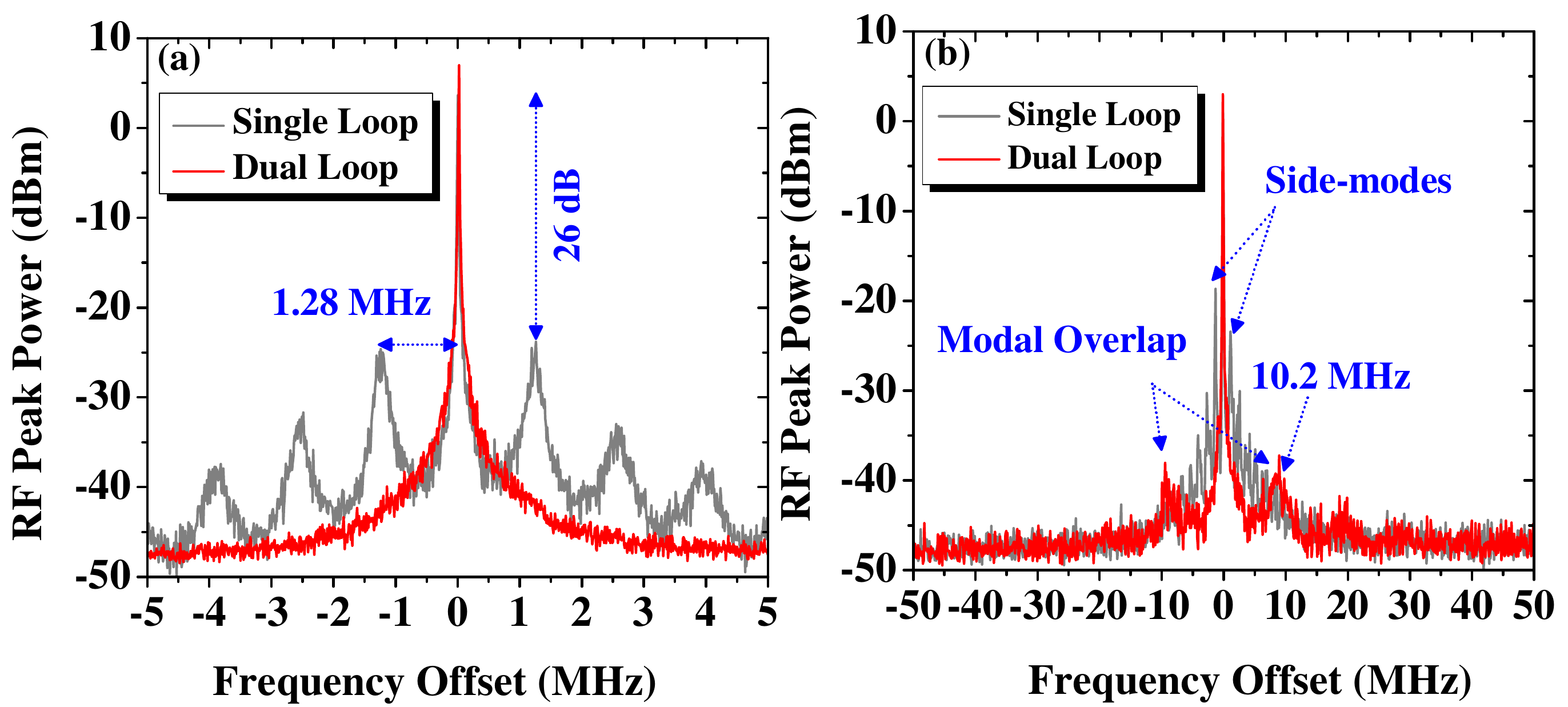}
\caption{RF Spectra of single-loop feedback with length 160m (gray line) and dual-loops having lengths 160m for loop-I and 20m for loop-II (red line) using frequency span (a) 10 MHz and (b) 100 MHz}
\label{fig:false-color}
\end{figure}
Measured phase noise traces for free-running condition (green line), single-loop (gray line) and dual-loop feedback with loop-II at 20m (red line), 53m (black line) and 80m (blue line) as functions of frequency offset from the fundamental mode-locked frequency are given in Fig. 5\par
\begin{figure}[htbp]
\centering
\includegraphics[width=\linewidth]{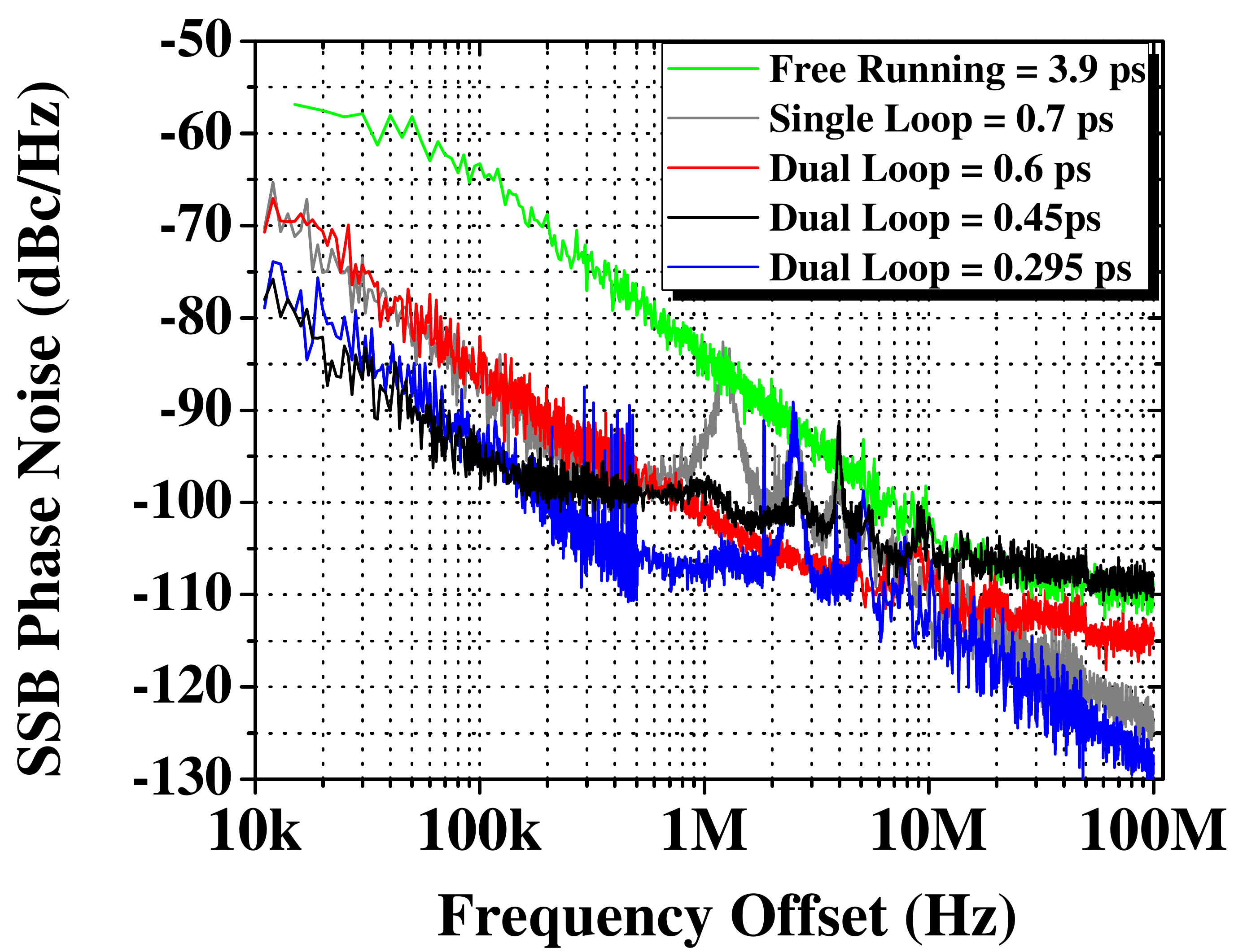}
\caption{SSB phase noise trace of free-running, single- and dual-loop feedback configurations with loop-I = 160m and loop-II 20m (red line), 53m (black line) and 80m (blue line) under fully resonant condition}
\label{fig:false-color}
\end{figure}
Comparison of RF linewidth and integrated timing jitter under stable resonant conditions, for three chosen lengths of second feedback cavity is shown in inset of Fig. 6. Measured integrated timing jitter in all dual-loop configurations was lower than for single-loop feedback. However, best suppression in external cavity side-modes was achieved with the second delay time $\sim$ 8x shorter than the first. Furthermore, the integrated timing jitter in this case was 16\% lower than for single-loop feedback. Reduction in timing jitter occurs due to suppression of external cavity side-modes relative to single-loop feedback. 
\begin{figure}[htbp]
\centering
\includegraphics[width=\linewidth]{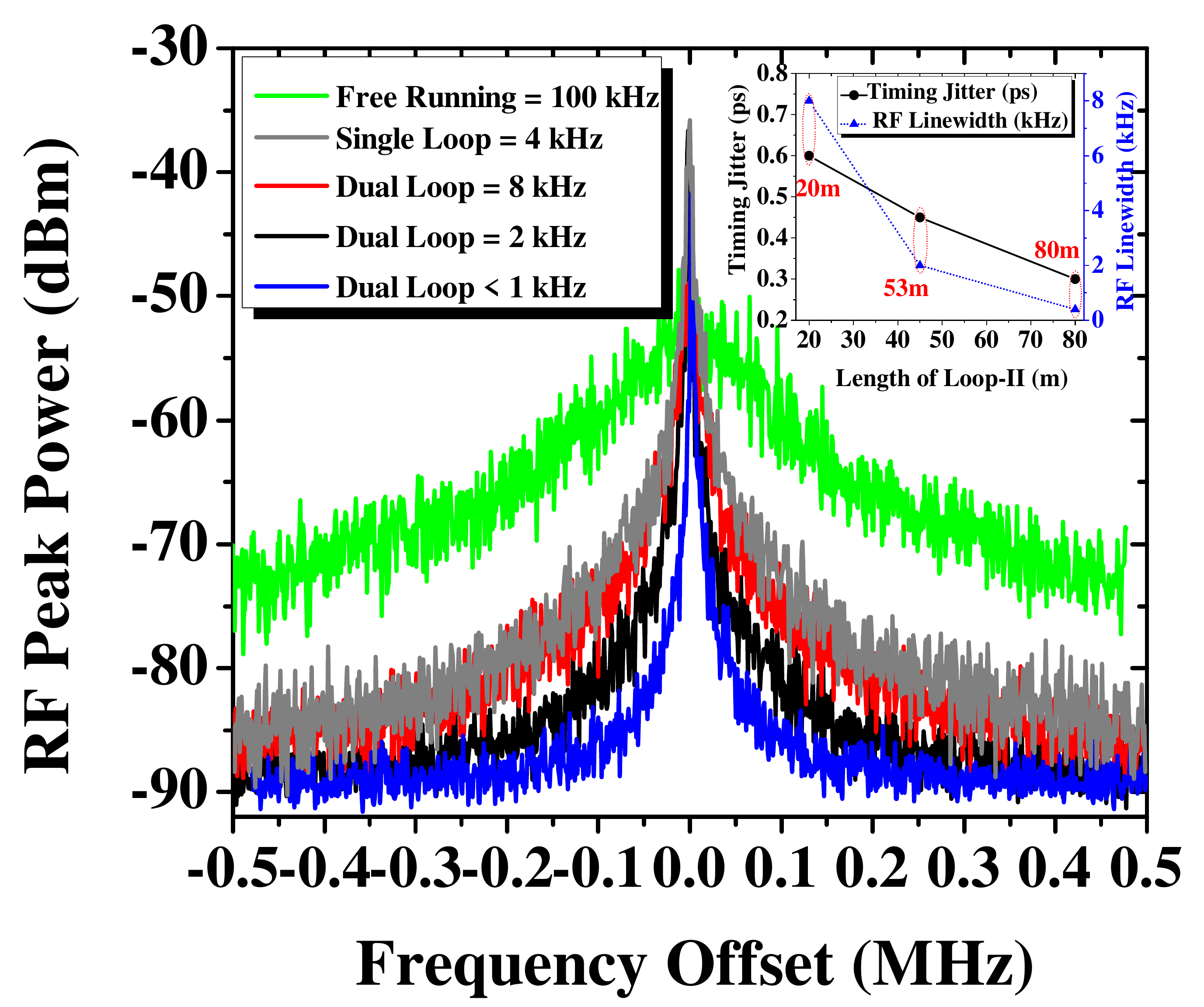}
\caption{
RF spectra for free-running, single- and dual-loop feedback with loop-I = 160m and loop-II 20m (red line), 53m (black line) and 80m (blue line); Inset shows RF Linewidth (blue triangles) and Integrated timing jitter (black circles) for dual-loop feedback with loop-I = 160m and loop-II 20, 53 and 80m }
\label{fig:false-color}
\end{figure}
In the literature [13, 14], it was experimentally observed that side-mode suppression was achieved when both feedback delays had a common multiple. This shows that effective suppression in external cavity side-modes is highly dependent on the length of the second loop. Recently, the influence of the second loop delay on suppression of external cavity side-modes [16] and timing jitter [17] was studied numerically. In this work, experimentally measured suppression of cavity side-modes and integrated timing jitter as a function of the second cavity delay correspond well with published numerical simulations [16, 17]. }\par
{In summary, an asymmetric dual-loop feedback method has been demonstrated which suppress additional noise resonances found in conventional single- and dual-loop feedback schemes. These results show that dual-loop feedback with precise alignment of the second loop delay effectively suppresses external cavity side modes and produces flat RF spectra closer to the main peak. Furthermore, by increasing the length of the second loop, significant reduction in RF linewidth and integrated timing jitter was produced. Our experimental results have validated recently published numerical simulations. Using this method, stable side-band-free integrated photonic oscillators based on mode-locked lasers may be developed which are feasible and attractive for many applications in optical telecommunications, time-domain multiplexing, frequency comb generation and as synchronized pulse sources or multi-wavelength lasers for wavelength-diversity or multiplexing.}\par
\textbf{Funding.} The authors acknowledge financial support from Science Foundation Ireland (grant 12/IP/1658) and the European Office of Aerospace Research and Development (grant FA9550-14-1-0204). 
  

\end{document}